\documentclass[a4paper,10pt,twoside]{cpc-hepnp}
\usepackage{CJK,upgreek,fancyhdr}
\usepackage{multicol}
\usepackage{graphicx}
\usepackage{booktabs}
\usepackage{amssymb,bm,mathrsfs,bbm,amscd}
\usepackage[tbtags]{amsmath}
\usepackage{lastpage}

\usepackage{color}
\usepackage[normalem]{ulem}

\begin{document}
\begin{CJK*}{GB}{gbsn}

\fancyhead[c]{\small Chinese Physics C~~~Vol. 41, No. 6 (2017) 065104}
\fancyfoot[C]{\small 065104-\thepage}

\footnotetext[0]{Received 28 November 2016}

\title{Scale-invariance in soft gamma repeaters\thanks{Supported by the National Natural Science Foundation of China (11375203, 11675182, 11690022 and 11603005), and the Fundamental Research Funds for the Central Universities (106112016CDJCR301206). }}

\author{%
      Zhe Chang $^{1,2}$
\quad Hai-Nan Lin $^{3}$ \\
\quad Yu Sang $^{1,2;1)}$\email{sangyu@ihep.ac.cn}
\quad Ping Wang $^{1}$
}
\maketitle

\address{%
$^1$Institute of High Energy Physics, Chinese Academy of Sciences, Beijing 100049, China\\
$^2$School of Physics, University of Chinese Academy of Sciences, Beijing 100049, China\\
$^3$Department of Physics, Chongqing University, Chongqing 401331, China\\
}

\begin{abstract}
The statistical properties of the soft gamma repeater SGR J1550--5418 are investigated carefully. We find that the cumulative distributions of fluence, peak flux and duration can be well fitted by a bent power law, while the cumulative distribution of waiting time follows a simple power law. In particular, the probability density functions of fluctuations of fluence, peak flux, and duration have a sharp peak and fat tails, which can be well fitted by a $q$-Gaussian function. The $q$ values keep approximately steady for different scale intervals, indicating a scale-invariant structure of soft gamma repeaters. Those results support that the origin of soft gamma repeaters is crustquakes of neutron stars with extremely strong magnetic fields.
\end{abstract}



\footnotetext[0]{\hspace*{-3mm}\raisebox{0.3ex}{$\scriptstyle\copyright$}2017
Chinese Physical Society and the Institute of High Energy Physics
of the Chinese Academy of Sciences and the Institute
of Modern Physics of the Chinese Academy of Sciences and IOP Publishing Ltd}%

\begin{multicols}{2}

\section{Introduction}\label{sec:introduction}

Soft gamma repeaters (SGRs) are high energy transients with persistent emissions of hard X-ray and soft $\gamma$-ray bursts. In 1979, a series of short and soft gamma-ray bursts from SGR 1806--20 were detected by the instruments aboard the \emph{Venera} spacecraft \cite{Mazets1979,Mazets1979b}. This was the first observation of SGRs. Since then, several SGRs have been discovered. When SGRs are quiescent, no burst is observed for many months or years. During this time, SGRs are likely to emit weak bursts below the thresholds of detectors. In their active periods, tens or hundreds of bursts are emitted, with quite a  soft spectrum compared to gamma-ray bursts (GRBs) \cite{Hurley2011}. Typically, the duration of a burst is about 100 ms and the waiting time (the time interval between two successive bursts) ranges from seconds to years \cite{Woods2004}. The high energy spectra ($E>25$ keV) of SGRs can be well fitted with an optically thin thermal bremsstrahlung (OTTB) model with temperature $20-40$ keV \cite{Gogus2001,Aptekar2001}. However, this model fails to fit the lower energy spectra. The bursts have peak luminosity up to $\thicksim10^{42}~\textrm{erg s}^{-1}$ and the observed fluences span the range from $10^{-10}$ to $10^{-4}$ erg cm$^{-2}$ \cite{Mereghetti2008}. The spin periods of SGRs are in the range  $2-12$ s, and increase with relatively large periods derivative $(\thicksim10^{-13}-10^{-10}~\textrm{s~s}^{-1})$. The rotational energy loss is not energetic enough to contribute to the total energy output. A widely accepted model for SGRs is the magnetar, that is, an isolated neutron star with extremely strong dipole magnetic fields of $\thicksim10^{14}-10^{15}$ G powers the repetitive gamma-ray bursts and persistent X-ray emissions \cite{Duncan1992,Kouveliotou1998,Kouveliotou1999,Thompson2002}. Another class of magnetar candidates are anomalous X-ray pulsars (AXPs), which share similar properties with SGRs in terms of their persistent X-ray emissions and SGR-like bursts \cite{Thompson1996,Gavriil2002,Hurley2011}.

SGR J1550--5418 was discovered by the \emph{Einstein} X-ray satellite and was originally named 1E 1547.0--5408 \cite{Lamb1981}. It was confirmed in the galactic plane survey in the \emph{Advanced Satellite for Cosmology and Astrophysics} \cite{Sugizaki2001}. Subsequent observations by \emph{XMM-Newton} and the \emph{Chandra X-Ray Observatory} identified the source as a magnetar candidate \cite{Gelfand2007}. Finally, radio observations provided crucial evidence to confirm the magnetar nature of the source. The surface magnetic dipole field of $2.2\times10^{14}$ G has been estimated from its observed spin period of 2.07 s and period derivative of $2.32\times10^{-11}~\textrm{s~s}^{-1}$ \cite{Camilo2007}. Due to the lack of bursting behavior, SGR J1550--5418 was initially classified as an AXP \cite{Camilo2007}. In 2008 the source began its burst-active episodes and hundreds of bursts have been observed by several instruments, including the \emph{Swift} Burst Alert Telescope (BAT) and \emph{Fermi} Gamma-ray Burst Monitor (GBM) \cite{Israe2010,Scholz2011,VonKienlin2012,VanDerHorst2012,Younes2014}. The source shares SGR-like behavior with SGRs 1806--20, 1900+14 and 1627--41. Therefore, it was reclassified as a SGR and renamed SGR J1550--5418 \cite{Palmer2009,Kouveliotou2009}. The source went through its three active episodes over half a year: the first in 2008 October, the second in 2009 January--February, and the last in 2009 March--April. In each active episode, 17, 352 and 15 bursts were observed by \emph{Fermi}/GBM, respectively \cite{Collazzi2015}. Von Kienlin et al. \cite{VonKienlin2012} performed temporal and spectral analysis for bursts from the first and last active episodes, and van der Horst et al. \cite{VanDerHorst2012} studied the second episode. Their results showed that the spectra of the first episode are best fitted with a single blackbody (BB) model, and the last episode with an optically thin thermal bremsstrahlung (OTTB) model. The spectra of the second episode are equally well fitted by an OTTB model, a Comptonized (COMP) model and two BB (BB+BB) model. Collazzi et al. \cite{Collazzi2015} provided the durations, spectral parameters, fluences, and peak fluxes for all the bursts of SGR J1550--5418.

One of the interesting properties of SGRs is the earthquake-like behavior, which implies that SGRs are likely to be self-organized criticality (SOC) systems. Cheng et al. \cite{Cheng1996} compared the statistical properties of the 111 bursts of SGR 1806--20 observed during 1979--1984 with the \emph{International Cometary Explorer (ICE)} satellite and earthquakes. They found that the cumulative energy distribution of bursts is well fitted by a power law with an index $\gamma=1.66$, similar to the well-known earthquake Gutenberg--Richter power law with an index $\gamma\thickapprox1.6$. Besides, bursts and earthquakes both have log-symmetric waiting time distributions. Prieskorn and Kaaret \cite{Prieskorn2012} studied the fluence distribution of over 3400 bursts from SGR 1806--20 and over 1963 bursts from SGR 1900+14 using the complete set of observations through 2011 March from the Proportional Counter Array (PCA) onboard the \emph{Rossi X-Ray Timing Explorer (RXTE)}, and showed that the cumulative event distribution can be fitted with a power law. G\"{o}\v{g}\"{u}\c{s} et al. \cite{Gogus1999,Gogus2000} studied the statistical properties of SGR 1900+14 and SGR 1806--20. SGR 1900+14 includes 187 bursts detected by the Burst and Transient Source Experiment (BATSE) aboard the \emph{Compton Gamma-Ray Observatory (CGRO)}, and 837 bursts detected by \emph{RXTE}/PCA. SGR 1806--20 includes 290 bursts detected with \emph{RXTE}/PCA, 111 bursts detected with \emph{CGRO}/BATSE and 134 bursts detected with \emph{ICE}. They showed that SGR 1900+14 and SGR 1806--20 shared earthquake-like statistical properties in terms of the distributions of fluence and waiting time, consistent with the results of Cheng et al. \cite{Cheng1996}. The earthquake-like behavior is a manifestation of SOC, supporting the idea that the energy origin of SGRs is starquakes of magnetars \cite{Duncan1992,Thompson1996}. Wang and Yu \cite{Wang2016} found the cumulative distributions of peak flux, fluence and duration of a repeating fast radio burst (FRB) show power law forms and proposed that FRBs may also be SOC events.

It was proposed that another SOC behavior of earthquakes is the scale-invariant structure of the energy fluctuations \cite{Wang2015}. The probability density functions (PDFs) of earthquake energy fluctuations at different times have fat tails with a $q$-Gaussian form \cite{Caruso2007}. Wang et al. \cite{Wang2015} showed that PDFs of energy fluctuations $E_{i+n}-E_i$ have a common function form at different scale intervals, i.e., the \emph{q} values are approximately equal for different $n$, which means that there is a scale-invariant structure in the energy fluctuations of earthquakes. It was also proposed that earthquakes of two faulting styles, i.e. the thrust and the normal, could be well explained by the Olami-Feder-Christensen (OFC) model, which is one of the most popular models of SOC.

An interesting question is whether SGRs share the same scale-invariant structure as earthquakes. In this paper, we study, as an example, the statistical properties of SGR J1550--5418. We calculate the cumulative distributions of the fluence, peak flux, duration, and waiting time in Section 2. In Section 3, for the first time, we calculate the probability distribution functions of fluctuations of these quantities. Finally, discussions and conclusions are given in Section 4.

\section{Statistical Properties of SGR J1550--5418}\label{sec:statistical}

In the five-year \emph{Fermi}/GBM magnetar burst catalog \cite{Collazzi2015}, a total of 384 bursts were observed in SGR J1550--5418, among which 354 bursts have measured fluence, 344 bursts have 4 ms peak flux, 382 bursts have duration, and 376 bursts have waiting time. All of the bursts were observed in the three active episodes in $2008-2009$. The duration is characterized by $T_{90}$, which is defined by the time that the cumulative counts rise from 5\% to 95\%. The waiting time is evaluated by the difference of $T_{90}$-starts of successive bursts. As SGR J1550-5418 consists of three isolated epochs, we discard the waiting time between the last burst of the first epoch and the first burst of the second epoch, and between the  last burst of the second epoch and the first burst of the third epoch. Figure \ref{fig:cumulative-distribution} shows the cumulative distribution of fluence, peak flux, duration and waiting time. The data is binned based on the Freedman-Diaconis rule \cite{Freedman1981}. The central value and $1\sigma$ error are evaluated by the average and the standard deviation of data points in each bin, respectively.

We first try to fit the cumulative distributions of fluence, peak flux, duration and waiting time with a simple power law,
\begin{equation}\label{eq:simple-power-law}
N(>x)= Ax^{\alpha}+B.
\end{equation}
We adopt the nonlinear least square fitting method to fit the binned data. The chi-squared is given by
\begin{equation}\label{eq:chi-squared}
\chi^2=\sum_i\frac{1}{{\sigma^2_i}}[N_i-N(x_i)]^2,
\end{equation}
where $(x_i, N_i)$ are the data points, $\sigma_i$ are the $1\sigma$ errors of data points. The best-fit parameters are derived using the Levenberg-Marquardt algorithm \cite{Seber2003,Bevington2003}.
However, we find that the cumulative distributions of fluence, peak flux and duration do not fit the simple power law well. The data points show an obvious excess at the right-hand end. Only the waiting time can be well fitted with Eq.~(\ref{eq:simple-power-law}), as shown in the lower right-hand panel of Figure \ref{fig:cumulative-distribution}. The best-fit parameter $\alpha$ of waiting time is $\alpha_W=-0.12\pm0.04$. We then use a bent power law
\begin{equation}\label{eq:bent-power-law}
N(>x)= A\left[1+\left(\frac{x}{x_b}\right)^{\alpha}\right]^{-1}
\end{equation}
to fit the cumulative distributions of fluence, peak flux and duration. The bent power law has a break at around $x_b$, below which the slope is approximately a constant, and above which it behaves as a simple power law. The bent power law was used to fit the power density spectra of gamma-ray bursts \cite{Guidorzi2016}. As shown in Figure \ref{fig:cumulative-distribution}, the cumulative distributions of fluence, peak flux and duration are well fitted with the bent power law. The best-fit parameters and the reduced chi-squared $\chi^2_{red}$ are listed in Table \ref{tab:fitting-parameter-powerlaw}.

\end{multicols}
\begin{center}
\includegraphics[width=8 cm]{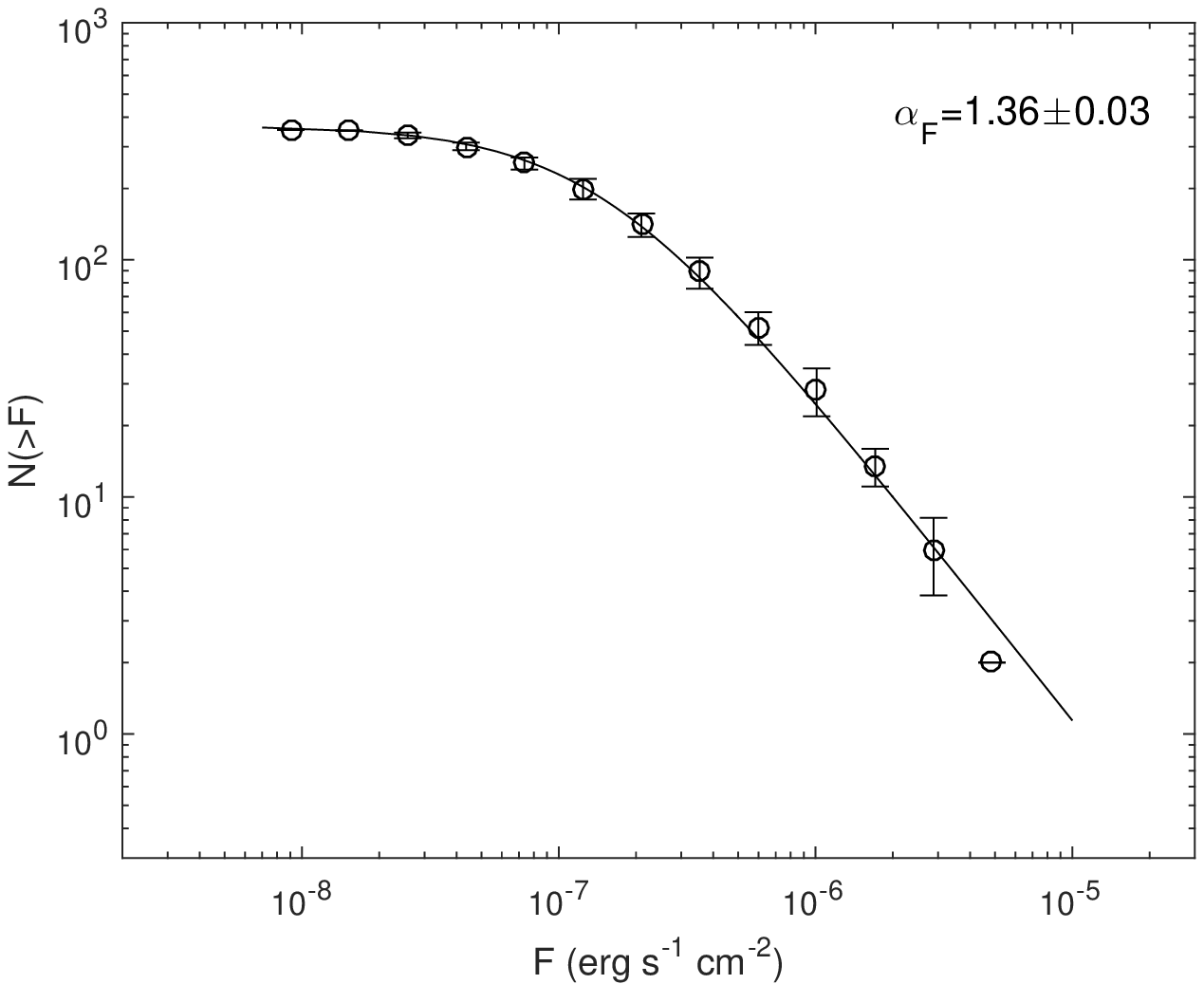}
\includegraphics[width=8 cm]{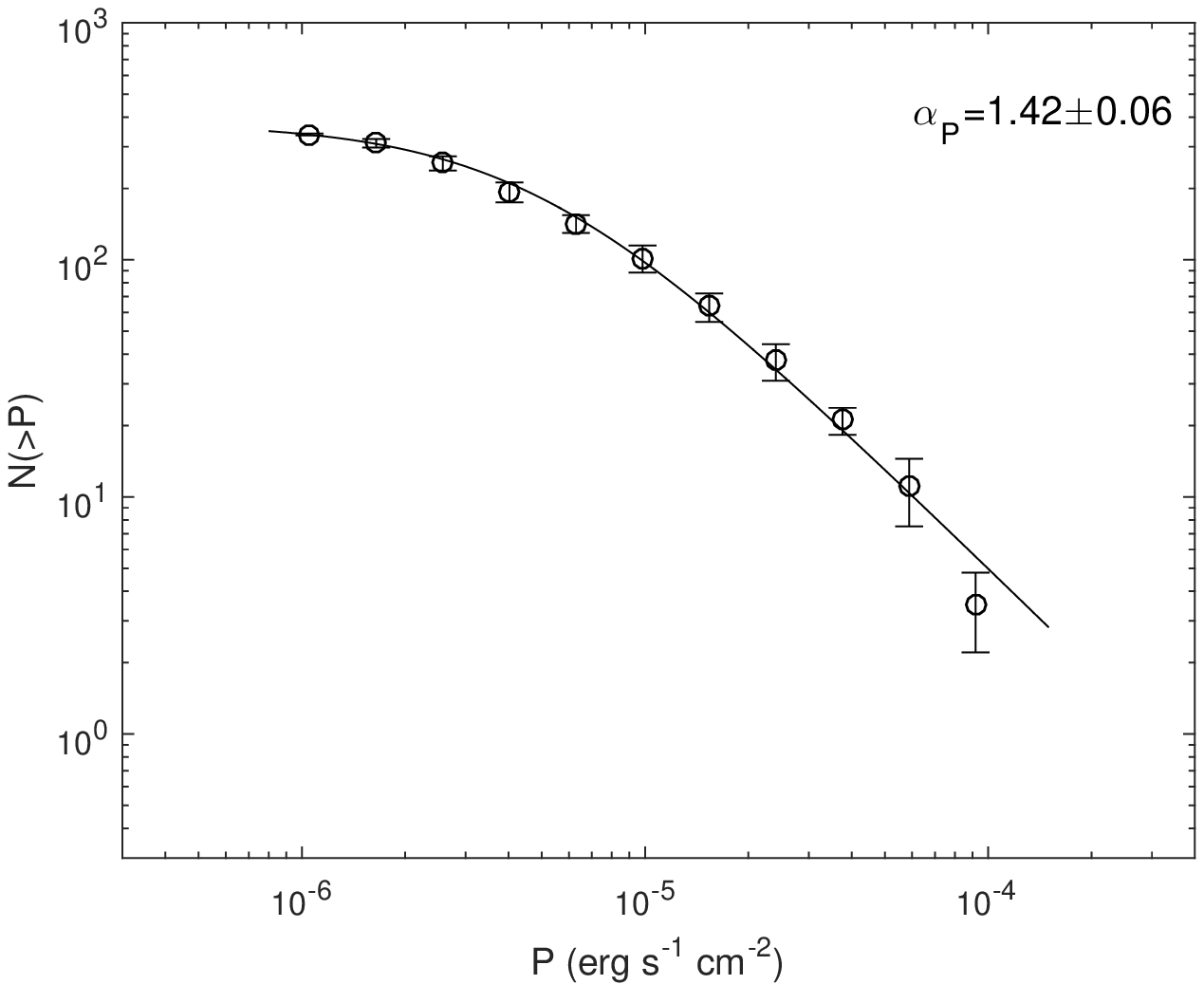}
\includegraphics[width=8 cm]{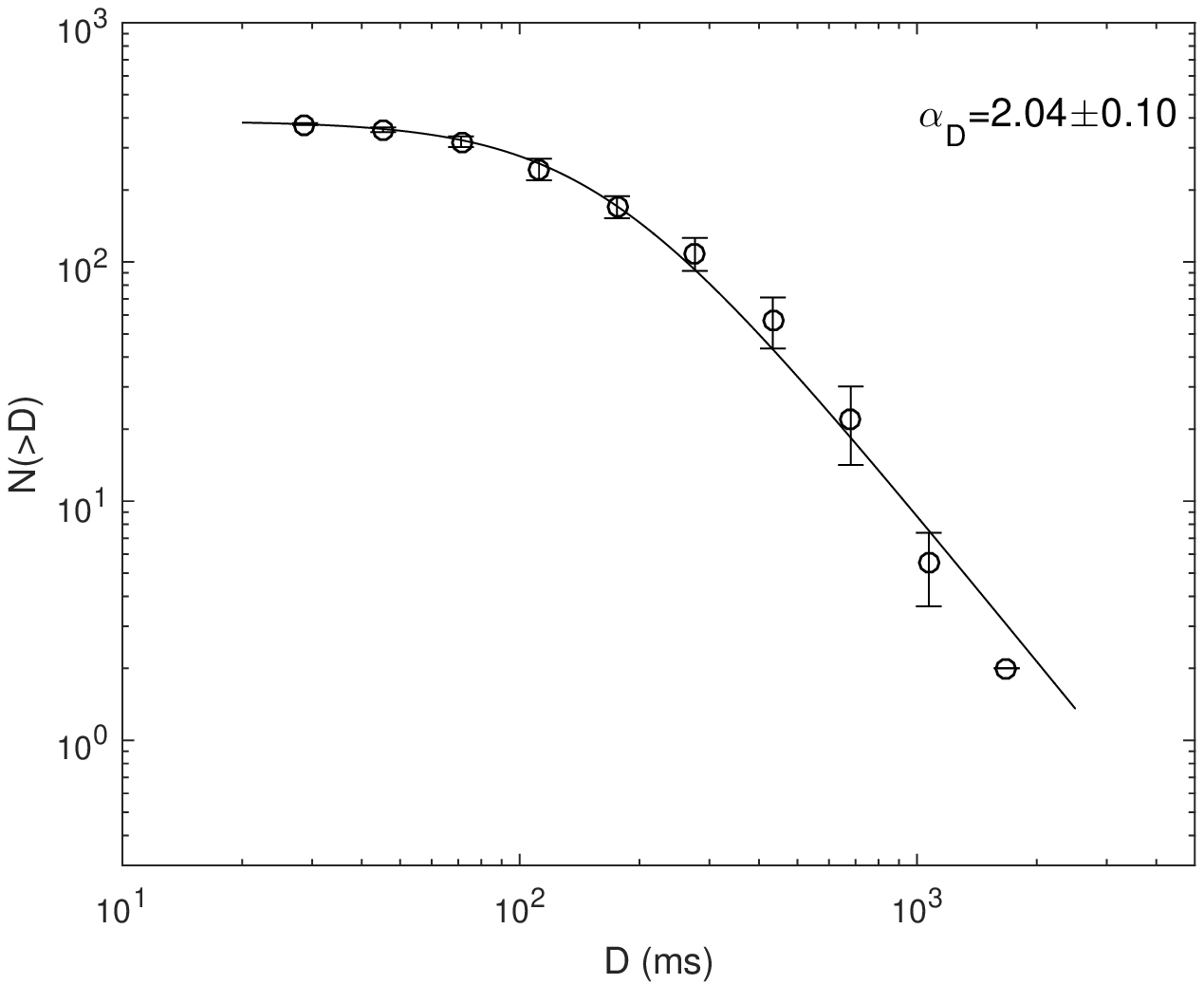}
\includegraphics[width=8 cm]{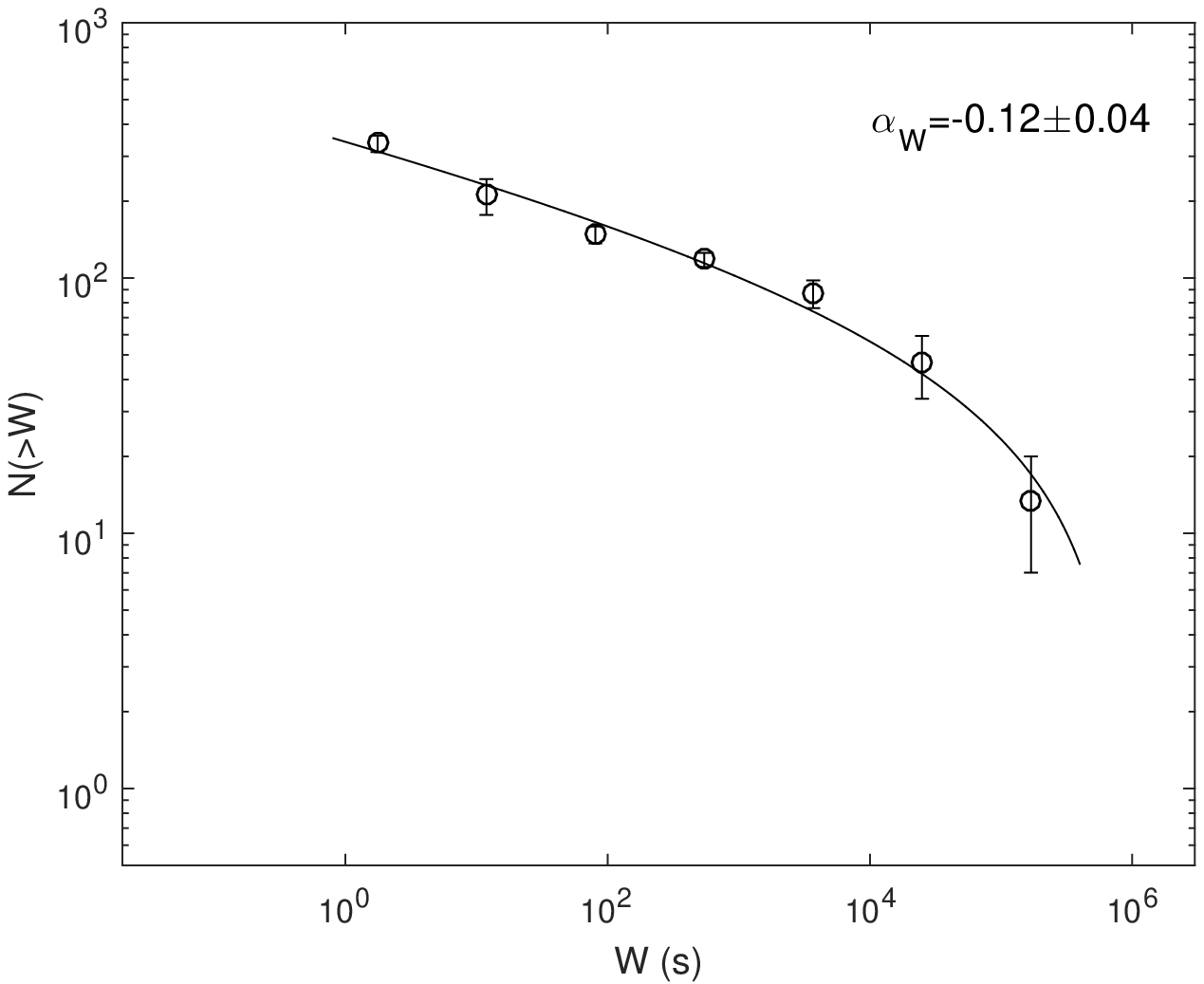}
\figcaption{\label{fig:cumulative-distribution}  The cumulative distributions of fluence (upper left panel), peak flux (upper right panel), duration (lower left panel) and waiting time (lower right panel). The solid lines are the best-fit results to bent power law (fluence, peak flux and duration) and simple power law (waiting time). The best-fit indices are $\alpha_F=1.36\pm0.03$, $\alpha_P=1.42\pm0.06$, $\alpha_D=2.04\pm0.10$, and $\alpha_W=-0.12\pm0.04$, respectively.}
\end{center}
\begin{multicols}{2}

\begin{center}
\tabcaption{ \label{tab:fitting-parameter-powerlaw} The best-fit power law indices $\alpha$ of the cumulative distributions of fluence, peak flux, duration and waiting time. Waiting time is fitted with the simple power law, while the remaining three parameters are fitted with the bent power law.}
\footnotesize
\begin{tabular*}{80mm}{c@{\extracolsep{\fill}}ccc}
\toprule Parameters & Values & $\chi^{2}_{red}$   \\
\hline
  $\alpha_F$ & $1.36\pm0.03$ & 0.23  \\
  $\alpha_P$ & $1.42\pm0.06$ & 0.67  \\
  $\alpha_D$ & $2.04\pm0.10$ & 0.65  \\
  $\alpha_W$ & $-0.12\pm0.04$ & 1.48   \\
\bottomrule
\end{tabular*}
\vspace{0mm}
\end{center}
\vspace{0mm}

\section{Probability density functions of fluctuations}\label{sec:fluctuation}

The fluence fluctuation of SGRs is defined as $Z_n =F_{i+n}-F_i$, where $F_i$ is the fluence of the $i$th burst in temporal order, and the integer $n$ denotes the temporal interval scale. Usually, $Z_n$ is rescaled by the standard deviation of $Z_n$, i.e. $z_n=Z_n/\sigma$, where $\sigma={\rm std}(Z_n)$. The fluctuations of peak flux, duration and waiting time are defined in a similar way. Figure \ref{fig:pdf} shows the PDFs of fluctuations of fluence, peak flux, duration and waiting time for $n$=1, 10 and 100. The data are binned based on the Freedman-Diaconis rule.
\end{multicols}
\begin{center}
\includegraphics[width=8 cm]{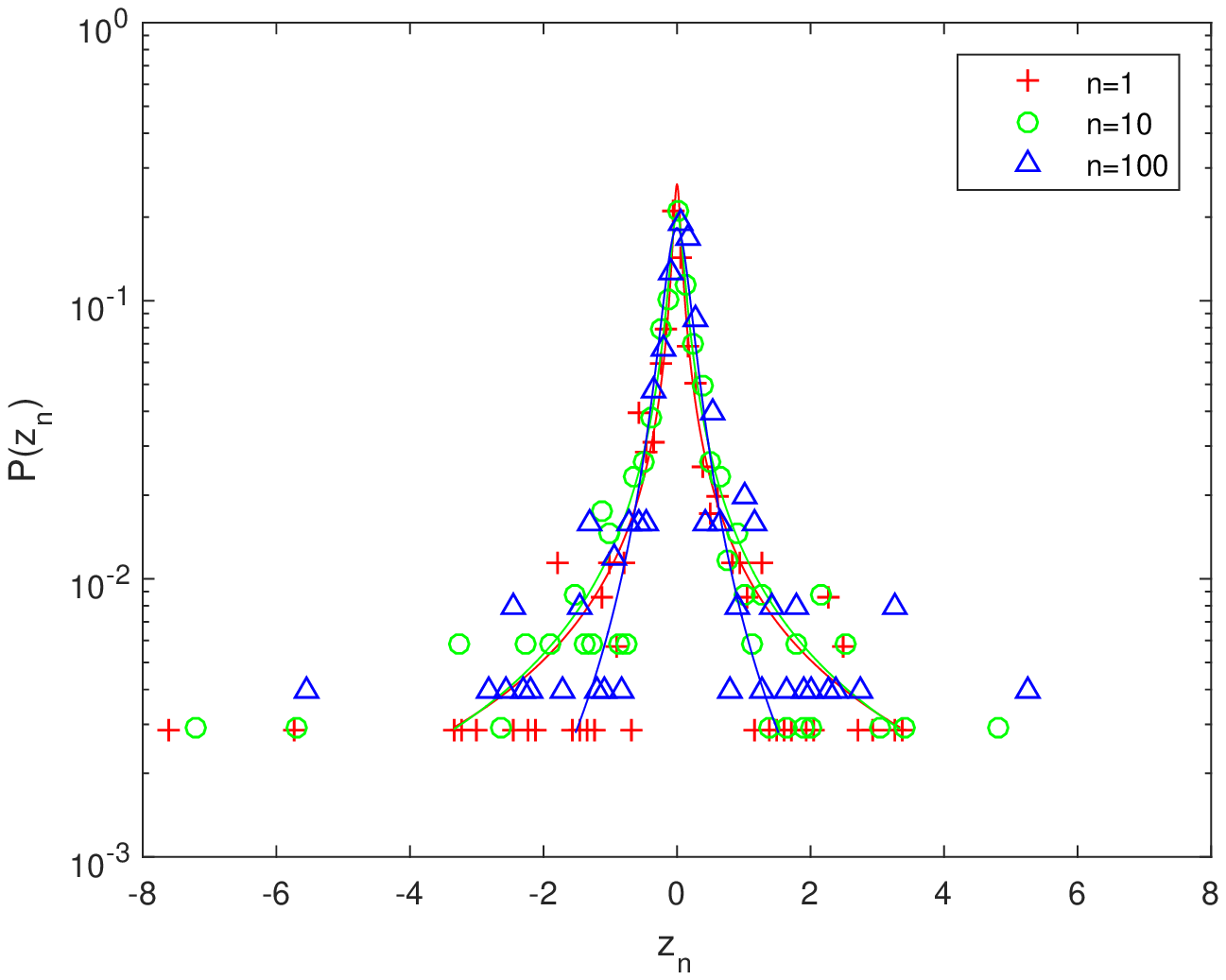}
\includegraphics[width=8 cm]{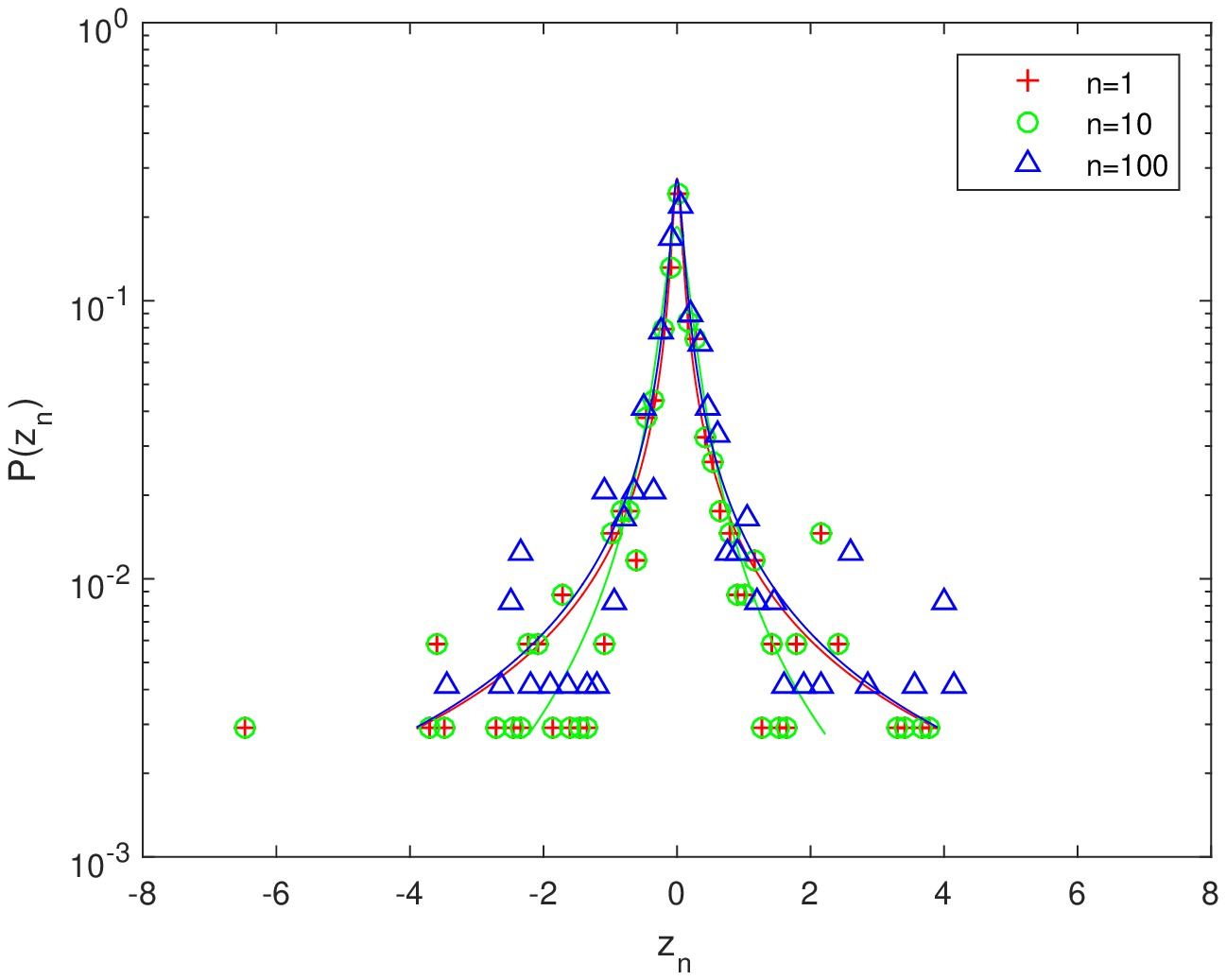}
\includegraphics[width=8 cm]{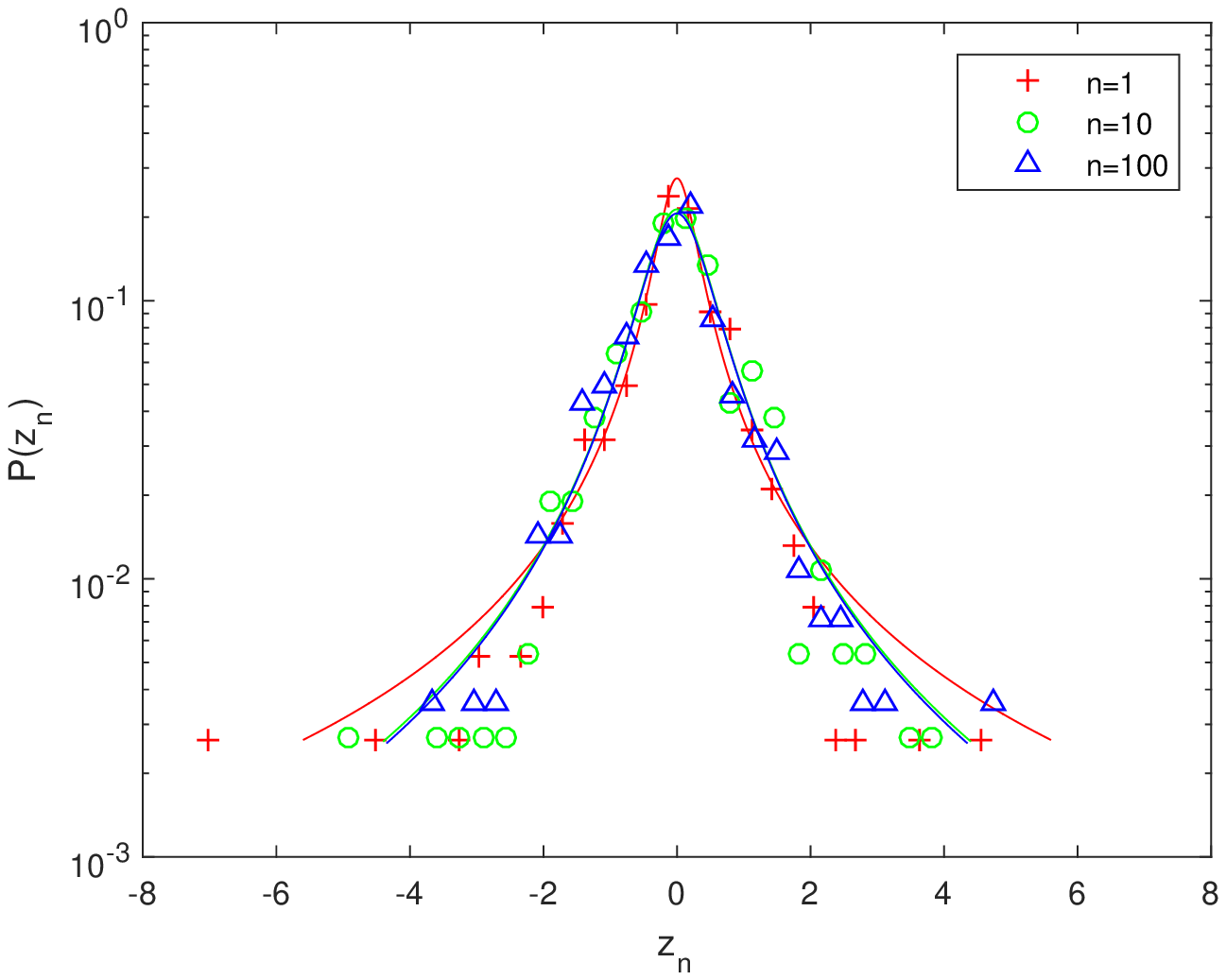}
\includegraphics[width=8 cm]{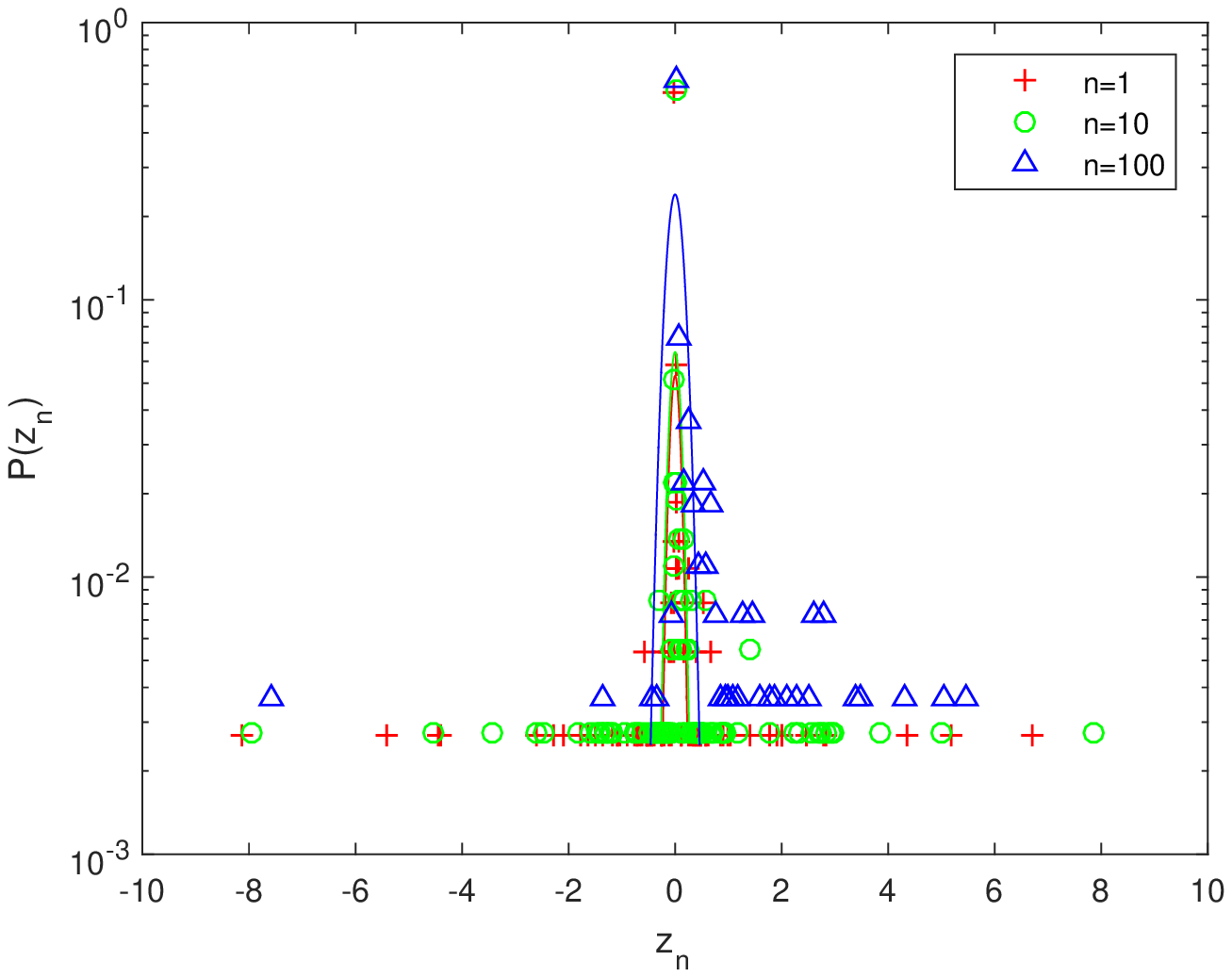}
\figcaption{\label{fig:pdf}(Color online) The PDFs of fluctuations of fluence (upper left panel), peak flux (upper right panel), duration (lower left panel) and waiting time (lower right panel) for $n=1$ (red crosses), $n=10$ (green circles) and $n=100$ (blue triangles). The solid lines are the best fits to a $q$-Gaussian distribution.}
\end{center}
\begin{multicols}{2}

From Figure \ref{fig:pdf}, we can see that $P(z_n)$ exhibits a sharp peak and fat tails like a Gaussian distribution, but the peak is much sharper than a Gaussian distribution. The fat tails mean that there are rare but large fluctuations, while the sharp peak means that small fluctuations are most likely to happen. The data points in Figure \ref{fig:pdf} are almost independent of $n$, indicating a common form of distribution. We use the $q$-Gaussian function
\begin{equation}\label{eq:qGaussian}
f(x)=\alpha[1-\beta(1-q)x^2]^{\frac{1}{1-q}}
\end{equation}
to fit $P(z_n)$, where $\alpha$, $\beta$ and $q$ are free parameters. The parameter $q$ denotes the deviation from a Gaussian distribution. The $q$-Gaussian distribution is a generalization of the Gaussian distribution and it reduces to a Gaussian distribution when $q\rightarrow1$. The $q$-Gaussian distribution exhibits heavy tails compared to a Gaussian distribution. In Figure \ref{fig:pdf}, the red, green and blue solid curves stand for the best-fit results for $n$=1, 10 and 100, respectively. The three curves are approximately superimposed near the peaks, and have some difference in the tails.

We use the nonlinear least squares fitting method to constrain the $q$ values. The best-fit parameters are listed in Table \ref{tab:fitting-parameter-qgauss}. The PDFs of fluctuations of fluence, peak flux and duration are well fitted with the $q$-Gaussian function, except for a few points at the fat tails. However, the $q$-Gaussian fit to waiting time shows that the $q$-values approach 1 regardless of $n$, implying that the fluctuation of waiting time is actually a Gaussian distribution.
\begin{center}
\tabcaption{ \label{tab:fitting-parameter-qgauss} The best-fit $q$ values in the $q$-Gaussian distribution for $n=1$, $10$ and $100$ }
\footnotesize
\begin{tabular*}{80mm}{c@{\extracolsep{\fill}}ccc}
\toprule Parameters & $n=1$ & $n=10$ & $n=100$  \\
\hline
  $q-\textrm{Fluence}$ & $2.84\pm0.10$ & $2.64\pm0.10$ & $1.90\pm0.25$ \\
  $q-\textrm{Peak Flux}$ & $2.82\pm0.10$ & $2.14\pm0.16$ & $2.71\pm0.14$ \\
  $q-\textrm{Duration}$ & $2.27\pm0.14$ & $1.94\pm0.16$ & $1.92\pm0.26$ \\
  $q-\textrm{Waiting Time}$ & $\sim1.00$ & $\sim1.00$ & $\sim1.00$ \\
\bottomrule
\end{tabular*}
\vspace{0mm}
\end{center}
\vspace{0mm}

Furthermore, we calculate the PDFs of fluctuation of fluence, peak flux and duration at different scale intervals $1\leq n\leq 100$, and fit the PDFs with the $q$-Gaussian function. We find that the $q$ values are approximately steady and independent of $n$, see Figure \ref{fig:scale-invariance}. For the fluence, the $q$ values are in the range of $1.50-3.20$, with a mean value of 2.41. For peak flux and duration, the mean $q$ values are 2.40 and 2.06, respectively. We also use the $q$-Gaussian function to fit the PDFs of fluctuation of waiting at scale intervals $1\leq n\leq 100$, and find that the best-fit $q$ values all approach $\sim1.00$. This means that waiting time has different behaviour.
\end{multicols}
\begin{center}
\includegraphics[width=12cm]{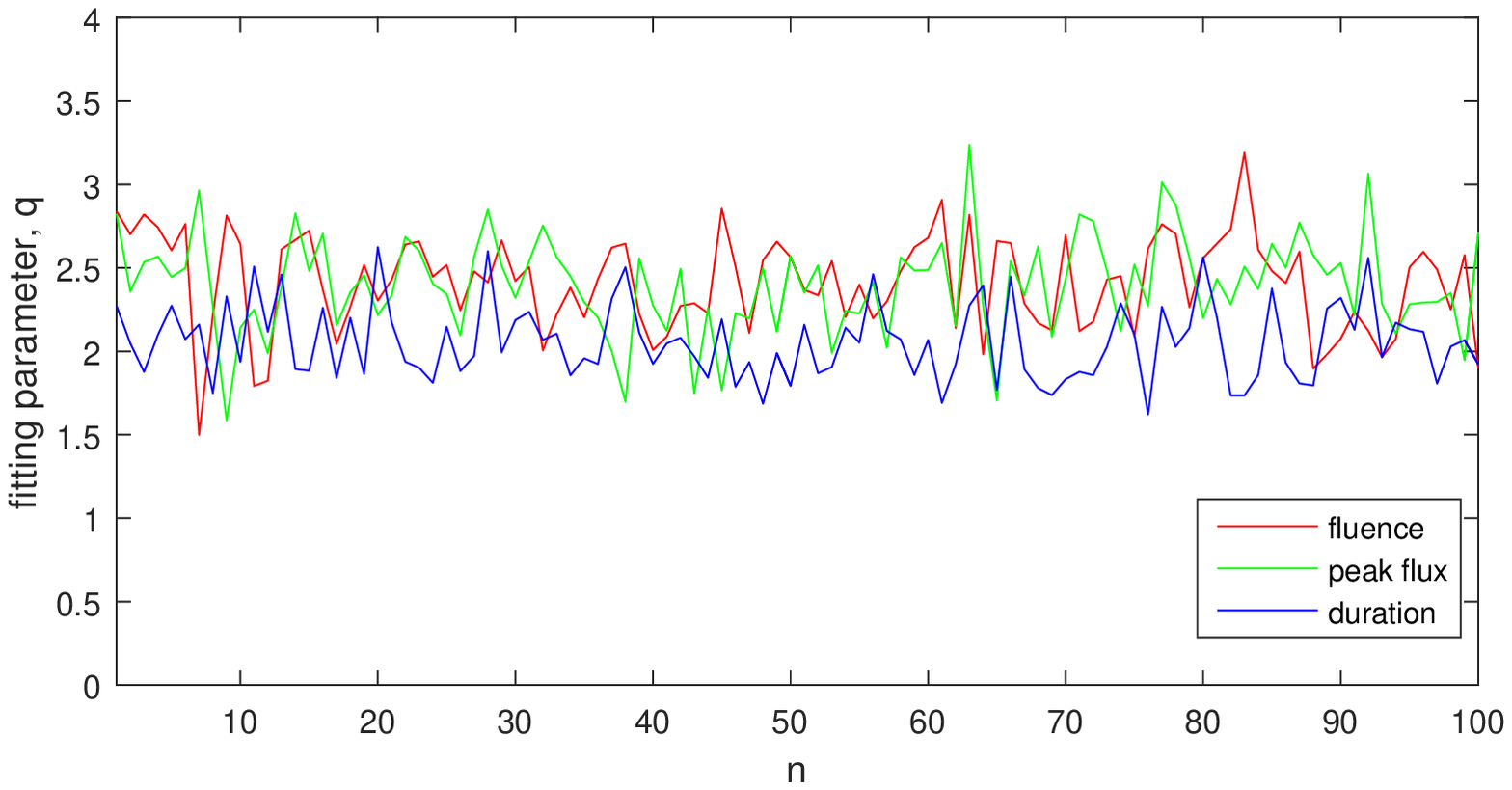}
\figcaption{\label{fig:scale-invariance}(Color online)  The best-fit $q$ values in the $q$-Gaussian distribution for $1\leq n\leq 100$.}
\end{center}
\begin{multicols}{2}

\begin{center}
\includegraphics[width=9 cm]{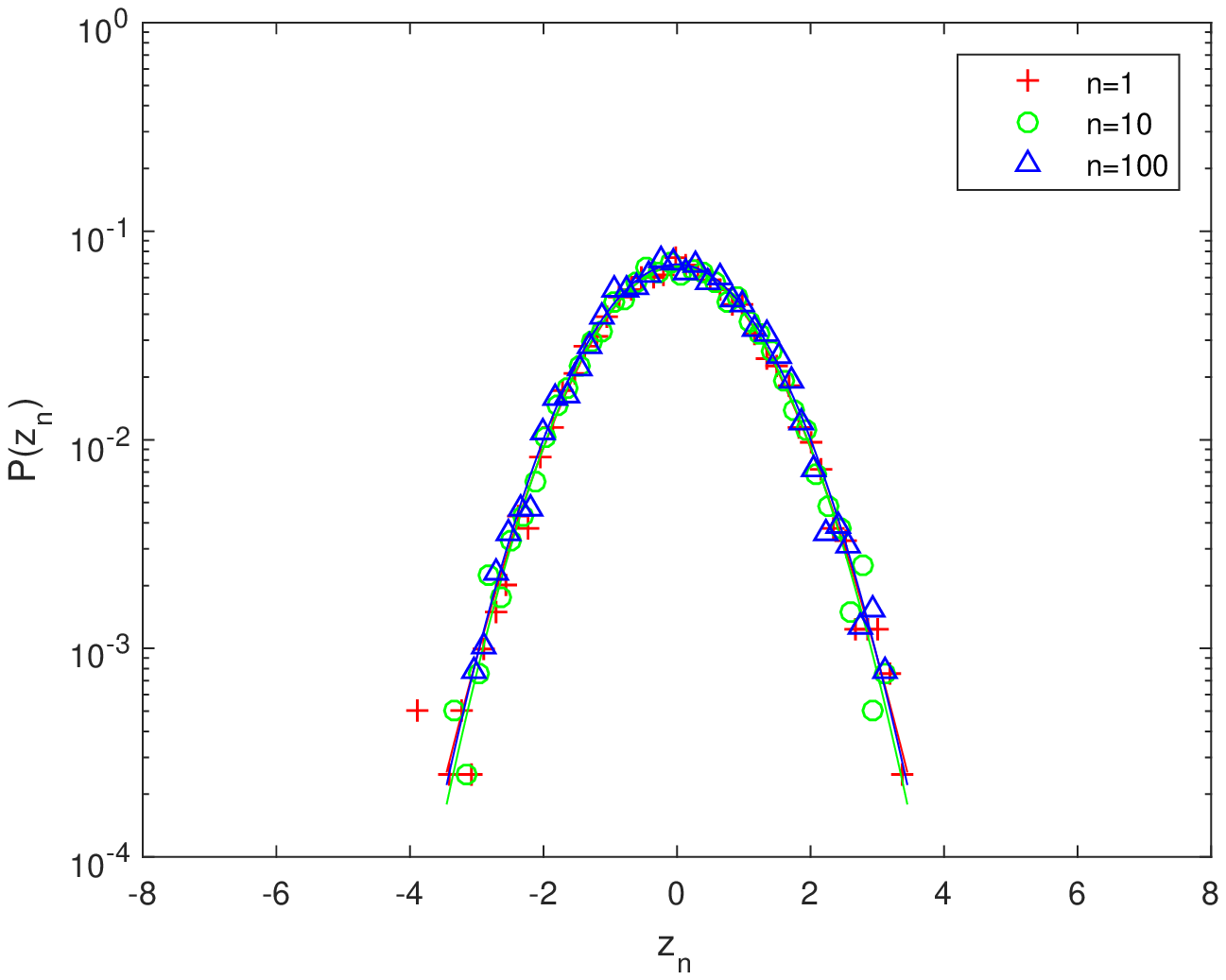}
\figcaption{\label{fig:pdf-simulation}(Color online)  The PDFs of fluctuations of mock fluence drawn from a Gaussian distribution for $n=1$ (red crosses), $n=10$ (green circles) and $n=100$ (blue triangles). The solid lines are the best fits to a $q$-Gaussian distribution with $q\sim1$ (i.e., Gaussian distribution). The solid lines for $n=1$, $10$ and $100$ almost overlap.}
\end{center}

To investigate whether the PDF of fluctuations in a $q$-Gaussian form can be generated from purely random bursts, we simulate 4000 bursts with fluence drawn from a Gaussian distribution with center $F_0=100$ and width of $F_0/4$. We calculate the fluctuations $Z_n$ in temporal order for $n$= 1, 10 and 100, rescale $Z_n$ to $z_n$ and bin the data using the Freedman-Diaconis rule. The $P(z_n)$ of $z_n$ is shown in Figure \ref{fig:pdf-simulation}. We fit the mock data with a $q$-Gaussian function, and find the mock data is well fitted with $q\approx 1$ for $n$= 1, 10 and 100, which means that the PDF of fluctuations of the mock data follow a Gaussian distribution, as we expected. The $P(z_n)$ in Figure \ref{fig:pdf-simulation} exhibits a smooth peak and steep tails compared to those in Figure \ref{fig:pdf}. For $n$= 1, 10 and 100, the best-fit curves nearly overlap, indicating a common Gaussian distribution. Moreover, we fit the $q$ value at different scale intervals $1\leq n\leq 100$, and find that the $q$ values are approximately steady with a mean of 1.03 in the range of 1.00--1.20, see Figure \ref{fig:scale-invariance-simulation}. This indicates that the fluctuation of mock data has a scale-invariant Gaussian distribution.
\end{multicols}
\begin{center}
\includegraphics[width=12cm]{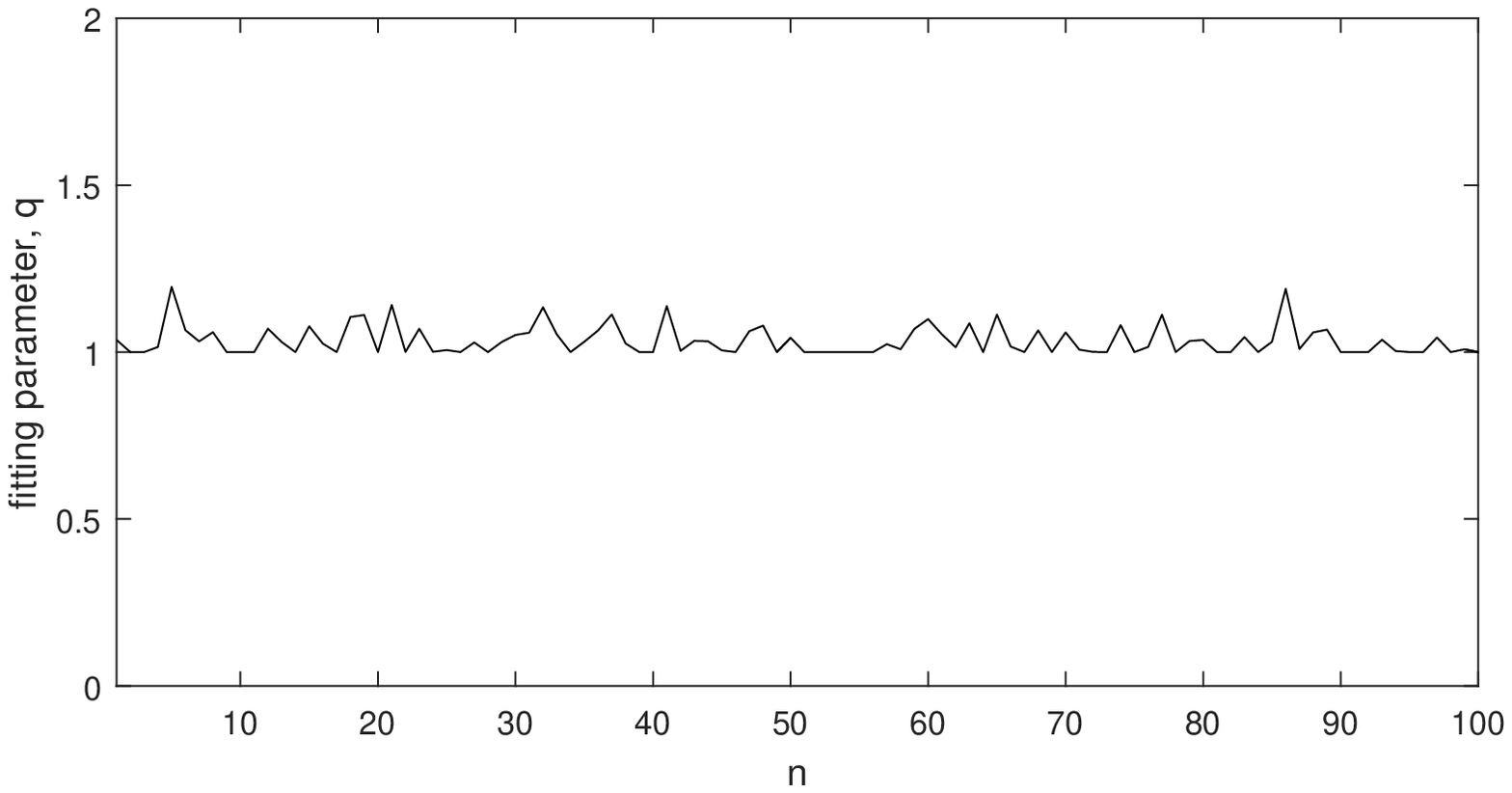}
\figcaption{The best-fit $q$ values of mock fluence in the $q$-Gaussian distribution for $1\leq n\leq 100$.}\label{fig:scale-invariance-simulation}
\end{center}
\begin{multicols}{2}

\section{Discussion and Conclusions}\label{sec:conclusion}

It was proposed that there is a scale-invariant structure in the energy fluctuations of earthquakes \cite{Wang2015}. The PDFs of earthquake energy fluctuations are well fitted with a $q$-Gaussian function, and the $q$ values are approximately equal at different time scales. In this paper, we investigated 384 bursts in the three active episodes of SGR J1550--5418 and found that the SGR shows similar behavior to earthquakes. The PDFs of fluence fluctuations were well fitted with a $q$-Gaussian function and the $q$ values keep approximately steady for different time scales. The earthquake-like behavior indicates that SGRs are likely to be SOC systems and supports the idea that the energy origin of SGRs is the starquakes of magnetars. The energy can be powered by the strain of solid crusts and strong magnetic fields of neutron stars. Besides, we showed that peak flux and duration share similar statistical properties with fluence, i.e. their PDFs of fluctuations also exhibit $q$-Gaussian distributions. However, the fluctuation of waiting time shows different behavior. The time gap between the three active episodes of SGR J1550-5418 and the discontinuous observations of the \emph{Fermi} satellite may explain the different behavior of the waiting time. Monte Carlo simulations show that $q$-Gaussian fluctuations could not arise from pure random bursts.

In summary, we have investigated the statistical properties of SGR J1550--5418. The cumulative distributions of fluence, peak flux and duration can be well fitted by a bent power law, and the PDFs of fluctuations of fluence, peak flux and duration can be well fitted by a $q$-Gaussian function. However, the waiting time shows different properties, which may be caused by the discontinuous observations of the \emph{Fermi} satellite. The PDFs exhibit a common functional form for different scale intervals, indicating a scale-invariant structure of SGRs. These properties are also common in earthquakes. This indicates that there may be some similarities between the energy origins of SGRs and earthquakes. The number of bursts in SGR J1550--5418 is not large enough to come to a definite conclusion; further statistical studies on a larger number of bursts are needed to verify these results.

\acknowledgments{We are grateful to X. Li, S. Wang and Z.-C. Zhao for useful discussions.}

\end{multicols}
\vspace{2mm}
\centerline{\rule{80mm}{0.1pt}}
\vspace{2mm}
\begin{multicols}{2}

\end{multicols}

\clearpage
\end{CJK*}
\end{document}